\documentclass[11pt,a4paper]{article}
\usepackage{jheppub}
\usepackage{amsmath,amsfonts,amssymb}
\usepackage[T1]{fontenc}
\usepackage[utf8]{inputenc}
\usepackage[british]{babel}
\usepackage{lmodern}
\usepackage{tikz}

\newcommand{\eq}[1]{\eqref{eq:#1}}
\newcommand{\fig}[1]{Fig.~\ref{fig:#1}}

\newcommand{\sect}[1]{Sec.~\ref{sec:#1}}

\toccontinuoustrue
\begin{document}
\author[a]{Tom Steudtner}
\affiliation[a]{
Department of Physics and Astronomy, U Sussex, Brighton, BN1 9QH, U.~K.
}
\emailAdd{T.Steudtner@sussex.ac.uk}
\title{General scalar renormalisation group equations at three-loop order}
\abstract{ For arbitrary scalar QFTs in four dimensions, renormalisation group equations of quartic and cubic interactions, mass terms, as well as field anomalous dimensions are computed at three-loop order in the $\overline{\text{MS}}$ scheme. Utilising pre-existing literature expressions for a specific model, loop integrals are avoided and templates for general theories are obtained. We reiterate known four-loop expressions, and from those derive $\beta$ functions for scalar masses and cubic interactions. As an example, the results are applied to compute all renormalisation group equations in $U(n) \times U(n)$ scalar theories.  }

%\keywords{RGEs, General Field Theories, Scalar, Three-loop, Dummy field method.}

\maketitle
\section{Introduction}
  The renormalisation group (RG) is a key instrument to connect and extrapolate physics to different scales as well as to study critical phenomena. 
  Hence, the computation of renormalisation group equations (RGEs) is a crucial issue in these kinds of studies, requiring high accuracies.
  In spite of the advent of non-perturbative methods, e.g. Wilsonian RG \cite{Wilson:1973jj,Polchinski:1983gv,Wetterich:1992yh,Morris:1993qb} or the gradient flow \cite{Luscher:2010iy,Luscher:2011bx}, perturbative RGEs have stood the test of time, due to  being systematic and extensible expansions that are reliable in the weak coupling regime.
  
  In the $\overline{\text{MS}}$ renormalisation scheme \cite{tHooft:1973mfk,Bardeen:1978yd}, a general framework is in place that allows one to extract RGEs for any renormalisable QFT without the need to perform loop calculations \cite{Machacek:1983tz,Machacek:1983fi,Machacek:1984zw}. This has provided universal access to all RGEs up to two-loop level \cite{Machacek:1983tz,Machacek:1983fi,Machacek:1984zw,Luo:2002ti,Schienbein:2018fsw, Sperling:2013eva, Sperling:2013xqa, Sartore:2020pkk} as well as complete three-loop gauge $\beta$ functions \cite{Pickering:2001aq, Mihaila:2012pz, Mihaila:2014caa, Poole:2019kcm}.  
  
  Due to its general applicability, it is desirable to extend this framework to higher loop levels. However, extensive calculations are required for this task, while the same computations in specific models are often comparatively easy.   
  Hence, it is not surprising that higher corrections only exist for theories of special interest. For instance, 5-loop results are available in simple gauge theories with fermions \cite{vanRitbergen:1997va, Czakon:2004bu, Herzog:2017ohr, Luthe:2017ttg}. Furthermore, RGEs of purely scalar theories have even been computed up to six loop orders for $O(n)$ \cite{Brezin:1974eb,Dittes:1977aq,Kazakov:1979ik,Chetyrkin:1981jq,Gorishnii:1983gp,Kazakov:1984km,Kleinert:1991rg,Batkovich:2016jus,Schnetz:2016fhy,Kompaniets:2017yct}, $O(n)\times O(m)$ \cite{Pelissetto:2001fi, Calabrese:2003ww, Kompaniets:2019xez} as well as cubic \cite{Nelson:1974xnq, Kleinert:1994td, Adzhemyan:2019gvv} symmetries. In four-dimensional Gross-Neveu-Yukawa and abelian Higgs models, the renormalisation group has been investigated up to four loops \cite{Mihaila:2017ble,Zerf:2017zqi,Zerf:2018csr,Ihrig:2019kfv,Zerf:2020mib}.
  
   All $\beta$ functions in the Standard Model (SM) are known up to three-loop order \cite{Mihaila:2012fm,Bednyakov:2012rb,Mihaila:2012pz,Bednyakov:2012en,Chetyrkin:2012rz,Bednyakov:2013eba,Chetyrkin:2013wya,Bednyakov:2013cpa,Bednyakov:2014pia} and even four-loop in the case of gauge couplings \cite{Bednyakov:2015ooa,Zoller:2015tha,Davies:2019onf} as well as QCD corrections to the Higgs self-interactions \cite{Martin:2015eia,Chetyrkin:2016ruf}.
  In the Two-Higgs-Doublet Model (THDM), three-loop $\beta$ functions have been determined for a type-III gauge-Yukawa sector \cite{Herren:2017uxn}, as well as the scalar potential \cite{Bednyakov:2018cmx}.

Foundations for advancing the general framework have been laid in \cite{Poole:2019kcm}, obtaining Weyl consistency conditions on four-loop gauge and three-loop Yukawa $\beta$ functions from the two-loop scalar quartic RGE. 
In contrast, no progress towards fully general three-loop quartic $\beta$ functions has been made. For a purely scalar potential however, RGEs have been determined in \cite{Jack:1990eb} and cross-checked in \cite{Jack:2018oec} using Weyl consistency conditions \cite{ Osborn:1991gm,Jack:2013sha,Poole:2019kcm}. Moreover, quartic four-loop $\beta$ functions and field anomalous dimensions have been obtained in \cite{Poole:2019kcm} using the same techniques. For long-range scalar theories, and using a different renormalisation scheme, three-loop RGEs have been recomputed in \cite{Benedetti:2020rrq}.

In this work, we demonstrate how to extract general three-loop $\beta$ functions and anomalous dimensions for an arbitrary scalar sector in the $\overline{\text{MS}}$ scheme. This will be achieved without using Weyl consistency conditions \cite{ Osborn:1991gm,Jack:2013sha,Poole:2019kcm} as a cross-check of various literature results. We also collect literature results at four loop order.
 In \sect{framework} we will briefly review formalisms and introduce notations. \sect{3-loop} will detail our approach to extract RGEs, and we present  general results in \sect{results}. For reference, we include expressions obtained in \cite{Jack:2018oec}, and compute four-loop $\beta$ functions for scalar masses and cubic interactions. 
 
 We will apply these expressions to compute RGEs for a scalar $U(n) \times U(n)$ matrix model, which has been of special interest e.g. due to the walking regime \cite{Holdom:1981rm,Kaplan:2009kr,Gorbenko:2018ncu,Benini:2019dfy,Antipin:2020rdw,Hansen:2017pwe} between its two complex fixed points \cite{Paterson:1980fc}, as well as in models with weakly-coupled asymptotic safety in pertubatively exact settings \cite{Litim:2014uca,Litim:2015iea,Bond:2017tbw,Bond:2017lnq,Bond:2019npq} and extensions of the SM \cite{Bond:2017wut,Hiller:2019tvg,Hiller:2019mou}.
 
\section{General Framework}\label{sec:framework}
Any perturbatively renormalisable QFT in four dimension can be embedded in the template Lagrangian \cite{Machacek:1983tz,Machacek:1983fi,Machacek:1984zw}
  \begin{equation}\label{eq:master-template}
  \begin{aligned}
    \mathcal{L} =
     & -\frac{1}{4} F_A^{\mu\nu} F^A_{\mu \nu} + \frac{1}{2} D^\mu \phi_a D_\mu \phi_a + i \psi_j^\dagger \sigma^\mu D_\mu \psi_j + \mathcal{L}_\text{gf} + \mathcal{L}_\text{gh}\\
     & - \frac{1}{2} \left(Y^a_{jk} \,\psi_j \varepsilon \psi_k \phi_a + Y^{a*}_{jk} \,\psi_j^\dagger \varepsilon \psi_k^\dagger \phi_a\right) - \frac{1}{4!} \lambda_{abcd} \,\phi_a \phi_b \phi_c \phi_d\,\\
     &  -\frac{1}{2} \left[\mathrm{m}_{jk}\, \psi_j \varepsilon \psi_k + \mathrm{m}^*_{jk} \,\psi^\dagger_j \varepsilon \psi^\dagger_k \right]
-\frac{m^2_{ab}}{2!} \phi_a \phi_b - \frac{h_{abc}}{3!} \phi_a \phi_b \phi_c\,,
  \end{aligned}
  \end{equation}
  which is formulated in terms of fermionic Weyl components $\psi_i$, real scalar fields $\phi_a$ and a generic gauge sector, including gauge-fixing and ghost terms $\mathcal{L}_\text{gf} + \mathcal{L}_\text{gh}$. Both scalar and fermionic indices $a,\,b,\,...$ and $i,\,j,\,...$ run over all field species and components, generation indices as well as gauge and flavour representations. All Yukawa couplings, scalar quartic and cubic interactions as well as fermion and scalar masses can hence be embedded into the respective tensor structures $Y^a_{ij}$, $\lambda_{abcd}$, $h_{abc}$, $\mathrm{m}_{ij}$ and $m^2_{ab}$. In particular, these quantities are chosen to be symmetric in all their scalar or fermionic indices, e.g. $\lambda_{abcd} = \lambda_{cabd}$.
  It is worth mentioning that the consistent treatment of left- and right-chiral fermions is non-trivial in dimensional regularisation. This is commonly known as the $\gamma_5$ problem, see \cite{Jegerlehner:2000dz} for a review, as well as \cite{Belusca-Maito:2020ala} for a more recent overview of works. As we are limiting ourselves to scalar theories, the issue is immaterial in this work, and does not affect any external results we utilise.

Using a dimensional regularisation in $d=4 - 2 \epsilon$ and the modified minimal subtraction scheme $(\overline{\text{MS}})$ \cite{Bollini:1972bi,Bollini:1972ui,tHooft:1973mfk,Bardeen:1978yd}, a multiplicative renormalisation procedure is established in the bare action \eq{master-template} via
\begin{equation}
  \phi_a^\text{(bare)} \mapsto \left[Z_{ab}^{1/2}(\mu) + \left(\delta Z\right)^{1/2}_{ab}\right] \phi_b , \qquad g_i^\text{(bare)} \mapsto \mu^{\epsilon \rho_i}\left[ g_i(\mu) + \delta g_i \right].
\end{equation}
Here, $\mu$ labels the renormalisation scale. Similar substitutions for fermion and gauge fields apply, introducing field strength renormalisation factors $Z$ and their corresponding counter-terms $\delta Z$. In the same manner, $g_i$ and $\delta g_i$ are placeholders for all couplings and their respective counter-terms. The numbers $\rho_i$ are determined by keeping the corresponding interaction operator $d$-dimensional after inserting the canonical dimensionality of the fields, e.g. $\rho=2$ for quartics and $\rho=1$ for gauge and Yukawa interactions. 
In minimal subtraction schemes, the counter-terms are independent of $\mu$ and can be expanded via
\begin{equation}
  \delta Z^{1/2}= \sum_{n=1}^\infty z_n \epsilon^{-n}, \qquad \qquad \qquad  \delta g_i =  \sum_{n=1}^\infty c_{i,n} \epsilon^{-n}.
\end{equation}
The overall scale independence of bare couplings and fields thus leads to a relation of the leading poles in the counter-terms to $\beta$ functions and field anomalous dimensions $\gamma_\phi$~\cite{Symanzik:1970rt, Callan:1970yg}, 
\begin{equation}
\begin{aligned}\label{eq:RGEs}
  \beta_{g_i} &= \frac{\partial g_i}{\partial \ln \mu} = - \epsilon \, \rho_i \,g_i - \rho_i \, c_{i,1} +  \sum_j \rho_j\, g_j \frac{\partial c_{i,1}}{g_j}, \\
  \gamma_\phi &= \frac{1}{2} \frac{\partial \ln Z}{ \partial \ln \mu} =  - \sum_i \rho_i \,g_i \frac{\partial z_1}{\partial g_i}
\end{aligned}
\end{equation}
which describe the renormalisation group flow of the renormalised couplings with respect to the scale $\mu$. 
Using perturbation theory, these quantities can be obtained in a loop expansion
\begin{equation}\label{eq:RGE-order}
  \beta_\lambda \big|_{\epsilon = 0} = \sum_{n=1} \frac{\beta^{\lambda,n\ell}}{(4\pi)^{2n}}, \qquad \qquad \qquad \gamma_\phi \big|_{\epsilon = 0} = \sum_{n=1} \frac{\gamma^{\phi,n\ell}}{(4\pi)^{2n}}
\end{equation} 
from the counter-terms of the same order. As those counter-terms of coupling constants are determined by the simple poles of the external leg contributions and proper vertex counter-terms, the $n$-loop order for each $\beta$-function contains two types of diagrams. There are tree-level  couplings contracted with $n$-loop anomalous dimensions, as well as $n$-loop proper vertex corrections, but not a mix of those diagrams. For instance, the $\beta$-function for scalar quartic couplings can be expressed as
\begin{equation} \label{eq:quartic-RGE}
  \beta^{\lambda,n\ell}_{abcd} = \Big(\sum_\text{perm.}\gamma^{\phi, n\ell} \cdot \lambda \Big)_{abcd} + V^{\phi^4,n \ell}_{abcd},
\end{equation}
where the first and second terms correspond to scalar leg corrections, and quartic vertex corrections, respectively.     
In combination with the general ansatz \eq{master-template}, all momentum integrals and spinor summations can be resolved, and the RGEs \eq{RGEs} can be expressed in terms of contracted generalised couplings. This provides a template to conveniently obtain RGEs for any renormalisable QFT by using the embedding into \eq{master-template}, without the need of loop calculations. This however comes at the cost of having to conduct an involved computation for the general theory \eq{master-template} once.

\section{Three-loop scalar contributions}\label{sec:3-loop}
In this section, we extract the pure scalar part of the field anomalous dimensions and quartic $\beta$-functions at three-loop order. To this end, we will assume the QFT given by the Lagrangian 
\begin{equation}\label{eq:ansatz}
  \mathcal{L} = \frac{1}{2} \partial^\mu \phi_a \partial_\mu \phi_a -\frac{1}{4!} \lambda_{abcd} \,\phi_a \phi_b \phi_c \phi_d.
\end{equation}
General $\beta$ functions for scalar masses and cubic interactions will be computed in the next section. For all these RGEs, gauge and fermionic interactions may be added without invalidating the results, such that they can be computed separately.
% Following \eq{RGE-order} and \eq{quartic-RGE}, we will list all scalar three-loop diagrams relevant for the computation of $\gamma^{\phi,3\ell}$ and $\beta^{\lambda,3\ell}$. In particular, the fact that momentum integrals of the shape
% \begin{equation}
%  \begin{tikzpicture}
%     \draw[black, thick] (2.em, -1.em) -- (4.em, 0) -- (6em, -1.em);
%     \draw[black, thick] (4.em, 1em) circle (1em);
%     \node at (4em, 0)[circle,fill,inner sep=.1em]{};
%      \node at (7em, 0)[]{$ \quad  =0$};
%  \end{tikzpicture}
% \end{equation}
% vanish in the $\overline{\text{MS}}$ scheme reduces the number of diagrams for the leg correction to 2, see \fig{leg}, as well as 8 vertex contributions, see \fig{vertex}. This set can be reduced further, e.g. it is immediately clear that $K^{(2)}$ in \fig{leg} does not yield corrections for the renormalised wave functions. However, we will retain such terms as a cross-check of our method.
Following \eq{RGE-order} and \eq{quartic-RGE}, all scalar three-loop Feynman diagrams relevant for the computation $\gamma^{\phi,3\ell}$ and $\beta^{\lambda,3\ell}$ can now be collected, as well as lower order ones with counter-term insertions. Using dimensional regularisation, the momentum integrations for each diagram can then be performed, while contractions of the couplings $\lambda_{abcd}$ in \eq{ansatz} are retained. Combining all contributions, the RGEs are extracted via \eq{RGEs}.

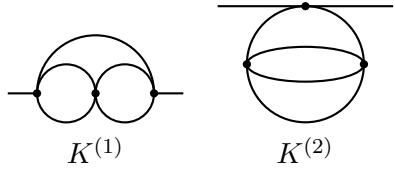
\begin{figure}
  \centering
  \begin{tabular}{cc}
    \begin{tikzpicture}
    \draw[black, thick] (1.em, 0.em) -- (2em, 0.em) arc (180:-0:2.em) -- (7em, 0.em);
    \draw[black, thick] (3em, 0.em) circle (1em);
    \draw[black, thick] (5em, 0.em) circle (1em);
    \node at (4em, 0)[circle,fill,inner sep=.1em]{};
    \node at (2em, 0)[circle,fill,inner sep=.1em]{};
    \node at (6em, 0)[circle,fill,inner sep=.1em]{};
    \end{tikzpicture} &
    \begin{tikzpicture}
    \draw[black, thick] (1.em, 0.em) -- (7em, 0.em);
    \draw[black, thick] (4.em, -2em) circle (2em);
    \draw[black, thick] (4.em, -2em) ellipse (2em and .6em);
    \node at (4em, 0)[circle,fill,inner sep=.1em]{};
    \node at (2em, -2em)[circle,fill,inner sep=.1em]{};
    \node at (6em, -2em)[circle,fill,inner sep=.1em]{};
    \end{tikzpicture}\\
    $K^{(1)}$ & $K^{(2)}$
  \end{tabular}
  \caption{Contractions of scalar quartic interactions in \eq{contr} that correspond to three-loop leg corrections.}
  \label{fig:leg}
\end{figure}

\begin{figure}
  \centering
  \begin{tabular}{cccc}
    \begin{tikzpicture}
      \draw[black, thick] (1em, 1em) -- (1em, -1em) -- (-1em, -1em) -- (-1em, 1em) --(1em, 1em);
      \draw[black, thick] (1.5em, 1.5em) -- (0.1em, 0.1em);
      \draw[black, thick] (-0.1em, -0.1em) -- (-1.5em, -1.5em);
      \draw[black, thick] (-1.5em, 1.5em) -- (1.5em, -1.5em);
      \node at (1em, 1em) [circle,fill,inner sep=.1em]{};
      \node at (-1em, 1em) [circle,fill,inner sep=.1em]{};
      \node at (1em, -1em) [circle,fill,inner sep=.1em]{};
      \node at (-1em, -1em) [circle,fill,inner sep=.1em]{};
    \end{tikzpicture}
    &
    \begin{tikzpicture}
    \draw[black, thick] (1.5em, 0.5em) -- (1.5em, -0.5em) -- (-1.5em, -0.5em) -- (-1.5em, 0.5em) --(1.5em, 0.5em);
    \draw[black, thick] (3em, 1.5em) -- (1.5em, 0.5em) .. controls (0.5em, 1.5em) and (-0.5em, 1.5em) .. ( -1.5em, 0.5em) -- (-3em,1.5em);
    \draw[black, thick] (3em, -1.5em) -- (1.5em, -0.5em) .. controls (0.5em, -1.5em) and (-0.5em, -1.5em) .. ( -1.5em, -0.5em) -- (-3em, -1.5em);
    \node at (1.5em, 0.5em)  [circle,fill,inner sep=.1em]{};
    \node at (-1.5em, 0.5em)  [circle,fill,inner sep=.1em]{};
    \node at (1.5em, -0.5em)  [circle,fill,inner sep=.1em]{};
    \node at (-1.5em, -0.5em)  [circle,fill,inner sep=.1em]{};
    \end{tikzpicture}
    &
    \begin{tikzpicture}
    \draw[black, thick] (-1.5em, 1.5em) -- (0,0) -- (-1.5em, -1.5em);
    \draw[black, thick] (1em, 0) circle (1em);
    \draw[black, thick] (3em, 0) circle (1em);
    \draw[black, thick] (5em, 0) circle (1em);
    \draw[black, thick] (7.5em, 1.5em) -- (6em,0) -- (7.5em, -1.5em);
    \node at (0, 0)[circle,fill,inner sep=.1em]{};
    \node at (2em, 0)[circle,fill,inner sep=.1em]{};
    \node at (4em, 0)[circle,fill,inner sep=.1em]{};
    \node at (6em, 0)[circle,fill,inner sep=.1em]{};
    \end{tikzpicture}
    &
    \begin{tikzpicture}
    \draw[black, thick] (-3em, 1.5em) -- (-1.5em,0) -- (-3em, -1.5em);
    \draw[black, thick] (3em, 1.5em) -- (1.5em,0) -- (3em, -1.5em);
    \draw[black, thick] (0, 0) circle (1.5em);
    \draw[black, thick] (0, 0) ellipse (0.3em and 1.5em);
    \node at (-1.5em,0)[circle,fill,inner sep=.1em]{};
    \node at (1.5em,0)[circle,fill,inner sep=.1em]{};
    \node at (0,-1.5em)[circle,fill,inner sep=.1em]{};
    \node at (0,1.5em)[circle,fill,inner sep=.1em]{};
    \end{tikzpicture}
    \\
    $T^{(1)}$ & $T^{(2)}$ & $T^{(3)}$ & $T^{(4)}$ \\
    \begin{tikzpicture}
    \draw[black, thick] (-3em, 1.5em) -- (-1.5em,0) -- (-3em, -1.5em);
    \draw[black, thick] (3em, 1.5em) -- (1.5em,0) -- (3em, -1.5em);
    \draw[black, thick] (0, 0) circle (1.5em);
    \draw[black, thick, fill= white] (0, -1.35em) ellipse (.8em and 0.5em);
    \draw[black, thick] (.77em,-1.35em) -- (-.77em,-1.35em);
    \node at (.77em,-1.32em)[circle,fill,inner sep=.1em]{};
    \node at (-.77em,-1.32em)[circle,fill,inner sep=.1em]{};
    \node at (-1.5em,0)[circle,fill,inner sep=.1em]{};
    \node at (1.5em,0)[circle,fill,inner sep=.1em]{};
    \end{tikzpicture}
    &
    \begin{tikzpicture}
    \draw[black, thick] (5em, 1.5em) -- (0.5em, -.75em);
    \draw[black, thick] (5em, -1.5em) -- (0.5em, .75em);
    \draw[black, thick] (4.1em, 0.5em) circle (0.5em);
    \draw[black, thick] (4.1em, -0.5em) circle (0.5em);
    \node at (2em, 0em)[circle,fill,inner sep=.1em]{};
    \node at (4.1em, 0em)[circle,fill,inner sep=.1em]{};
    \node at (4.1em, 1em)[circle,fill,inner sep=.1em]{};
    \node at (4.1em, -1em)[circle,fill,inner sep=.1em]{};
    \end{tikzpicture}
    &
    \begin{tikzpicture}
    \draw[black, thick] (5em, 1.5em) -- (2em, 0em);
    \draw[black, thick] (5em, -1.5em) -- (2em, 0em);
    \draw[black, thick] (-0.5em, -.75em) -- (1em, 0em) -- (-0.5em, .75em);
    \draw[black, thick] (4.1em, 0em) ellipse (0.5em and 1em);
    \draw[black, thick] (1.5em, 0) circle (0.5em);
    \node at (2em, 0em)[circle,fill,inner sep=.1em]{};
    \node at (1em, 0em)[circle,fill,inner sep=.1em]{};
    \node at (4.1em, 1em)[circle,fill,inner sep=.1em]{};
    \node at (4.1em, -1em)[circle,fill,inner sep=.1em]{};
    \end{tikzpicture}
    &
    \begin{tikzpicture}
    \draw[black, thick] (-1.5em, -.75em) -- (0em, 0em) -- (-1.5em, .75em);
    \draw[black, thick] (1.5em, .75em) -- (1.5em, -.75em) -- (0em, 0em) -- (1.5em, .75em) -- (3em, 1.5em);
    \draw[black, thick] (1.5em, -.75em) -- (3em, 0) -- (1.5em, .75em);
    \draw[black, thick] (4.em, .5em) -- (3em, 0) to [bend left = 60] (1.5em, -.75em);
    \node at (0em, 0em)[circle,fill,inner sep=.1em]{};
    \node at (1.5em, .75em)[circle,fill,inner sep=.1em]{};
    \node at (1.5em, -.75em)[circle,fill,inner sep=.1em]{};
    \node at (3em, 0em)[circle,fill,inner sep=.1em]{};
    \end{tikzpicture}
    \\
    $T^{(5)}$ & $T^{(6)}$ & $T^{(7)}$ & $T^{(8)}$
  \end{tabular}
  \caption{Contractions of scalar interactions in \eq{contr} corresponding to three-loop quartic vertex contributions.}
  \label{fig:vertex}
\end{figure}
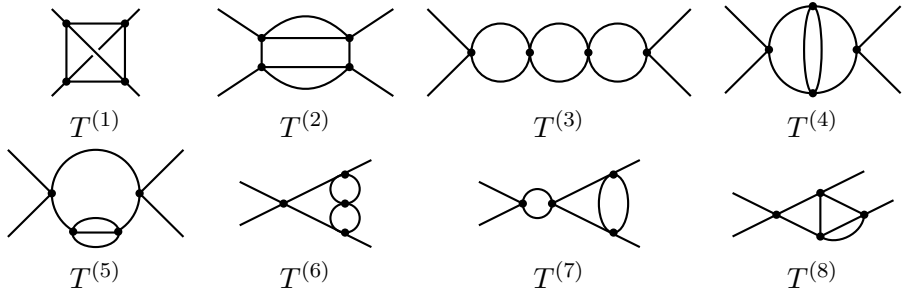

However, we will employ a different strategy: instead of dealing with diagrams directly, we provide an ansatz for $\gamma^{\phi,3\ell}$ and $\beta^{\lambda,3\ell}$ consisting of all possible contractions of generalised quartic couplings from \eq{ansatz}. Diagrammatic contributions after momentum integrations and counter-term insertions are then absorbed into an unknown prefactor for each of the contractions. These are denoted as $\kappa_{1,2}$ and $\tau_{1..8}$ in our ansatz
%The next step would be to compute the momentum integrals for these graphs as well as lower order diagrams with counter-term insertions. However, we will employ a different strategy: we will reinterpret \fig{leg} and \fig{vertex} as index contractions for the generalised quartic couplings of \eq{ansatz} in the RGEs $\gamma^{\phi,3\ell}$ and $\beta^{\lambda,3\ell}$ \eq{RGE-order} instead of diagrams. This is possible because both momentum integrals and counter-term insertions only constitute unknown prefactors $\kappa_{1,2}$ and $\tau_{1..8}$ in the general ansatz
  \begin{equation}\label{eq:RGEs-3L}
  \begin{aligned}
    \gamma^{\phi,3\ell}_{ab} &= \sum_{i=1}^2 \kappa_i \, K^{(i)}_{ab},\\
    \beta^{\lambda,3\ell}_{abcd} &= \sum_{i=1}^2 \kappa_i \left(K^{(i)}_{ae} \lambda_{ebcd} + K^{(i)}_{be} \lambda_{aecd} + K^{(i)}_{ce} \lambda_{abed} + K^{(i)}_{de} \lambda_{abce} \right)+ \sum_{j=1}^8 \tau_j \, T^{(j)}_{abcd} \,, 
  \end{aligned}
  \end{equation}
  where we have introduced the contractions 
    \begin{equation}\label{eq:contr}
  \begin{aligned}
  K^{(1)}_{ab} &=  \lambda_{acde} \lambda_{bcfg} \lambda_{defg}, &K^{(2)}_{ab} &=  \lambda_{abcd} \lambda_{cefg} \lambda_{defg}, \\
    T^{(1)}_{abcd} &=  \lambda_{aefg} \lambda_{behi} \lambda_{cfhj} \lambda_{dgij} ,  \quad &T^{(2)}_{abcd} &= \lambda_{aefg} \lambda_{fghb} \lambda_{ceij} \lambda_{ijhd} + \text{5 perm.},\\
    T^{(3)}_{abcd} &= \lambda_{abef} \lambda_{efgh} \lambda_{ghij} \lambda_{ijcd} + \text{2 perm.}, &T^{(4)}_{abcd} &= \lambda_{abef} \lambda_{egij} \lambda_{fhij} \lambda_{ghcd} + \text{2 perm.},\\
    T^{(5)}_{abcd} &= \lambda_{abef} \lambda_{ehij} \lambda_{hijg} \lambda_{fgcd} + \text{2 perm.}, &T^{(6)}_{abcd} &=\lambda_{abef} \lambda_{cegh} \lambda_{ghij} \lambda_{dfij} + \text{5 perm.},\\
    T^{(7)}_{abcd} &= \lambda_{abef} \lambda_{efgh} \lambda_{egij} \lambda_{dhij} + \text{5 perm.}, &T^{(8)}_{abcd} &=\lambda_{abef}  \lambda_{cegh} \lambda_{fgij} \lambda_{dhij} + \text{11 perm.}\,,
  \end{aligned}
\end{equation}
  displayed in \fig{leg} and \fig{vertex}. Terms containing index summations
\begin{equation}
 \begin{tikzpicture}
    \draw[black, thick] (2.em, -1.em) -- (4.em, 0) -- (6em, -1.em);
    \draw[black, thick] (4.em, 1em) circle (1em);
    \node at (4em, 0)[circle,fill,inner sep=.1em]{};
     \node at (7em, 0)[]{$ \quad  =0$};
 \end{tikzpicture}
\end{equation}
  have been excluded since the momentum integrals of corresponding diagrams vanish in dimensional regularisation, yielding $2+8$ potential contributions to leg and vertex renormalisation. Similar arguments can be employed to reduce the ansatz \eq{contr} further, e.g. removing $K^{(2)}$, but we will retain such terms as a cross-check of our method.
   Permutations in \eq{contr} ensure the overall symmetry of $\beta^\lambda_{abcd}$ and therefore $\lambda_{abcd}$. The number in front of `$\text{perm.}$' accounts for the exchanges of external indices that produce inequivalent expressions, e.g. $\lambda_{abef} \lambda_{efcd}$ is equivalent to $\lambda_{abef} \lambda_{efdc}$, but not to $\lambda_{acef} \lambda_{efbd}$.
  In \cite{Benedetti:2020rrq}, the contractions $T^{(1..8)}$ in \eq{contr} as well as the corresponding diagrams have been obtained as well.
  
   
  Here we point out that the 10 open parameters $\kappa_i$ and $\tau_j$ may be fixed by comparing the ansatz \eq{RGEs-3L} against literature results that are available for specific models. Hence, this approach allows us to obtain the most general RGEs by extrapolation of pre-existing results, without the need for explicit loop computations. This is possible because the number of unknown prefactors is relatively small. 
  
  Suitable three-loop computations are available for the SM Higgs sector \cite{Bednyakov:2013eba,Chetyrkin:2012rz,Chetyrkin:2013wya,Bednyakov:2013cpa}, theories with $O(n)$, $O(n)\times O(m)$ and cubic scalars \cite{Brezin:1974eb,Dittes:1977aq,Kazakov:1979ik,Chetyrkin:1981jq,Gorishnii:1983gp,Kazakov:1984km,Kleinert:1991rg,Batkovich:2016jus,Schnetz:2016fhy,Kompaniets:2017yct,Pelissetto:2001fi,Calabrese:2003ww,Kompaniets:2019xez,Nelson:1974xnq,Kleinert:1994td,Adzhemyan:2019gvv}, and the scalar potential of the  Two-Higgs-Doublet-Model (THDM) \cite{Bednyakov:2018cmx}.
  The results for the SM Higgs \cite{Bednyakov:2013eba,Chetyrkin:2012rz,Chetyrkin:2013wya,Bednyakov:2013cpa} do not provide enough data to fix all coefficients. However, neglecting all other interactions, the complex doublet can actually be described by a $O(4)$ real scalar. The general case of theories with $n$ scalars and a $O(n)$ symmetry is given by the Lagrangian
\begin{equation}\label{eq:O(n)}
  \mathcal{L} = \tfrac12 \partial_\mu \vec{\phi} \cdot \partial^\mu \vec{\phi} - \tfrac12 m^2 \vec{\phi} \cdot \vec{\phi} - \tfrac14 \lambda \left(\vec{\phi} \cdot \vec{\phi} \right)^2.
\end{equation}    
The respective three-loop RGEs \cite{Brezin:1974eb} can be matched against the ansatz \eq{RGEs-3L}, where each power of $n$ in the RGEs $\beta^{\lambda,3\ell}/\lambda^4$, $ \gamma^{m^2,3\ell}/\lambda^3$ and $\gamma^{\phi,3\ell}/\lambda^3$ gives a separate condition on the $\kappa_i$ and $\tau_j$. Unfortunately, the data extracted by this ansatz is insufficient for resolving all $\tau_i$. Alternatively, we will select a subset of the THDM, featuring two vector-like, complex doublet scalars $\Phi_{1,2}$.
The quartic potential  
\begin{equation}
  V = \tfrac12\lambda_1 \left(\Phi_1^\dagger \Phi_1^{\phantom{\dagger}}\right)^2 + \tfrac12\lambda_2 \left(\Phi_2^\dagger \Phi_2^{\phantom{\dagger}}\right)^2 + \lambda_3 \left(\Phi_1^\dagger \Phi_1^{\phantom{\dagger}}\right) \left(\Phi_2^\dagger \Phi_2^{\phantom{\dagger}}\right)
\end{equation}
is protected by an $U(2)_{\Phi_1} \times U(2)_{\Phi_2}$ symmetry, where each scalar has its own subgroup. This allows for the permutation $\Phi_1 \leftrightarrow \Phi_2$ and hence $\lambda_1 \leftrightarrow \lambda_2$. 
Translated into our own notation, the three-loop $\beta$ functions computed in \cite{Bednyakov:2018cmx} 
\begin{equation}
  \begin{aligned}\label{eq:THDM-RGE}
  \beta^{3\ell}_{\lambda_{1,2}} &=  3[299 + 168 \zeta(3)] \lambda_{1,2}^4 + 102\,\lambda_{1,2}^2 \lambda_3^2 - 78\,\lambda_1 \lambda_2 \lambda_3^2 + 4[107+ 48 \zeta(3)] \lambda_{1,2} \lambda_3^3 \\
  &\phantom{=} + 9\,\lambda_{2,1}^2 \lambda_3^2 + 192\,\lambda_{2,1} \lambda_3^3 + 6[1 + 8 \zeta(3)] \lambda_3^4, \\
  \beta^{3\ell}_{\lambda_{3\phantom{,3}}} &= 2[49 + 12 \zeta(3)] \lambda_3^4 + 36[7 + 4 \zeta(3)](\lambda_1+ \lambda_2) \lambda_3^3 + \tfrac92[29 + 48 \zeta(3)] (\lambda_1^2 + \lambda_2^2) \lambda_3^2 \\
  &\phantom{=} + 180 \,\lambda_1 \lambda_2 \lambda_3^2 + \tfrac{513}{2} (\lambda_1^3+\lambda_2^3)\lambda_3
  \end{aligned}
\end{equation}
allow us to extract the complete set of coefficients from \eq{RGEs-3L} which yields the definite solution
\begin{equation}\label{eq:solution}
  \begin{aligned}
    \kappa_1 &=  -1/16, \qquad &\kappa_2 &= 0,\qquad  &\tau_1 &= 12 \zeta(3), \qquad &\tau_2 &= -1/2, \qquad &\tau_3 &= 0, \\
     \tau_4 &= 1/2, \qquad &\tau_5 &= -3/8, \qquad & \tau_6 &= -1/2, \qquad & \tau_7 &= 0, \qquad & \tau_8 &= 2,
  \end{aligned}
\end{equation}
featuring the zeta function $\zeta(3) \approx 1.202$. We find that the contractions $K^{(2)}$, $T^{(3)}$ and $T^{(7)}$ do not give corrections to the RGEs. In fact, momentum integrals vanish for certain choices of external momenta in those three diagrams. This eliminates their contributions in the $\overline{\text{MS}}$ scheme, which is in accordance with the calculation conducted in \cite{Brezin:1974eb}, where $K^{(2)}$, $T^{(3)}$ and $T^{(7)}$ have been neglected.
Curiously, in $O(n)$ models like \eq{O(n)}, the fact that $\tau_3 = 0$ is the reason why leading-$n$ contributions $ n^3 \lambda^4 $ are absent in $\beta_\lambda^{3\ell}$, see \cite{Brezin:1974eb}. 

  This procedure is extendable to higher loop orders. At four loops, results with $O(n)$ vector, $O(n)\times O(m)$ matrix or cubic scalars \cite{Brezin:1974eb,Dittes:1977aq,Kazakov:1979ik,Chetyrkin:1981jq,Gorishnii:1983gp,Kazakov:1984km,Kleinert:1991rg,Batkovich:2016jus,Schnetz:2016fhy,Kompaniets:2017yct,Pelissetto:2001fi,Calabrese:2003ww,Kompaniets:2019xez,Nelson:1974xnq,Kleinert:1994td,Adzhemyan:2019gvv} are available for matching. 
  Several other four-loop calculations are found in the literature, also featuring gauge and Yukawa interactions, for instance \cite{Zerf:2020mib,Ihrig:2019kfv,Zerf:2018csr,Zerf:2017zqi}. 
  However, projecting out the scalar sectors reduces these theories to the $O(n)$ case.

  In \cite{Jack:2018oec}, 4+19 non-vanishing tensor structures have been found for the field anomalous dimensions and quartic vertex corrections. We agree with these findings, but are unable to determine all 23 open parameters from matching against the literature expressions.  In \cite{Jack:2018oec} these gaps were filled by utilising Weyl consistency conditions, and the full four-loop results were obtained. Correcting the number of permutations for the graph $g^\lambda_{4s}$ (in the notation of \cite{Jack:2018oec}), we will list the obtained expressions in the next section.

\section{Discussion} \label{sec:results}
In summary, we obtain the general result for the purely scalar part of any three- and four-loop scalar field anomalous dimension, in accordance with \cite{Jack:1990eb,Jack:2018oec} 
\begin{equation}\label{eq:WF-gen}
\begin{aligned}
    \gamma_{ab}^{\phi,3\ell} &= - \tfrac{1}{16} \lambda_{acde} \lambda_{defg} \lambda_{fgcb},\\
    \gamma_{ab}^{\phi,4\ell} &= - \tfrac{5}{64} \lambda_{acde} \lambda_{defg} \lambda_{fghi} \lambda_{hicb} - \tfrac5{96} \lambda_{acde} \lambda_{efgh} \lambda_{fghi} \lambda_{icdb} \\
    &\ \phantom{=}  + \tfrac{13}{96}\lambda_{acde} \lambda_{dfgh} \lambda_{efgi} \lambda_{chib} + \tfrac13 \lambda_{acde} \lambda_{cdfg} \lambda_{efhi} \lambda_{ghib}.
\end{aligned}
\end{equation}
All contributions to $\gamma^{\phi,3\ell}$ and $\gamma^{\phi,4\ell}$ are symmetric in the external indices, which means the expressions cover all external leg corrections in $\beta^{\lambda,3\ell}$ and $\beta^{\lambda,4\ell}$ respectively.
The RGEs of the quartic interactions $\lambda_{abcd}$ read
\begin{equation}\label{eq:quartic-gen}
  \begin{aligned}
    \beta_{abcd}^{\lambda, 3\ell} &= \left[\gamma_{ae}^{\phi,3\ell} \lambda_{ebcd} + \gamma_{be}^{\phi,3\ell} \lambda_{aecd} + \gamma_{ce}^{\phi,3\ell} \lambda_{abed} + \gamma_{de}^{\phi,3\ell} \lambda_{abce}\right] \\
    &\phantom{=} + 12 \zeta(3) \,\lambda_{aefg} \lambda_{behi} \lambda_{cfhj} \lambda_{dgij} 
     - \tfrac12 \left[ \lambda_{aefg} \lambda_{bfgh} \lambda_{ceij} \lambda_{dhij} + \text{5 perm.}\right] \\
     &\phantom{=} + \tfrac12 \left[ \lambda_{abef} \lambda_{egij} \lambda_{fhij} \lambda_{cdgh} + \text{2 perm.}\right] 
    - \tfrac38 \left[ \lambda_{abef} \lambda_{ehij} \lambda_{ghij} \lambda_{cdfg} + \text{2 perm.}\right]\\
    &\phantom{=} - \tfrac12 \left[ \lambda_{abef} \lambda_{cegh} \lambda_{ghij} \lambda_{dfij} + \text{5 perm.}\right] + 2 \left[ \lambda_{abef}  \lambda_{cegh} \lambda_{fgij} \lambda_{dhij} + \text{11 perm.}\right],\\[0.5em]
    \beta_{abcd}^{\lambda, 4\ell} &= \left[\gamma_{ae}^{\phi,4\ell} \lambda_{ebcd} + \gamma_{be}^{\phi,4\ell} \lambda_{aecd} + \gamma_{ce}^{\phi,4\ell} \lambda_{abed} + \gamma_{de}^{\phi,4\ell} \lambda_{abce}\right] \\
    &\phantom{=} + \left[\lambda_{abef} \lambda_{cdgh} \bigcirc_{ef|gh}  +  \text{ 5 perm.} \right]  + \left[\lambda_{abef} \lambda_{cghi} \lambda_{djkl} \bigtriangleup_{ef|ghi|jkl}  +  \text{ 11 perm.} \right] \\
     &\phantom{=} + \left[ \lambda_{aefg} \lambda_{bhij} \lambda_{cklm} \lambda_{dnpq} \,\square_{efg|hij|klm|npq} + \text{ 23 perm.} \right].
  \end{aligned}
\end{equation}
Here we have introduced tensor structures that encode the non-external part of the four-loop diagrams
\begin{equation}
  \begin{aligned}
  \bigcirc_{ab|cd} &= \tfrac12\left(1-\zeta(3)\right)\lambda_{abef} \lambda_{efgh} \lambda_{cdgh}  + \tfrac7{24} \delta_{ac} \lambda_{befg} \lambda_{fghi} \lambda_{dehi} \\
  &\phantom{=} + \left(\zeta(3) - \tfrac{11}6\right)\lambda_{acef} \lambda_{begh} \lambda_{dfgh} , \\
  \bigtriangleup_{ab|cde|fgh} &= 
  \tfrac5{12} \lambda_{adef} \lambda_{bcgh} + \tfrac23 \delta_{ac} \lambda_{bfgi} \lambda_{dehi}  + \left(\tfrac5{12} - \tfrac12 \zeta(3)\right) \delta_{cf} \lambda_{adei} \lambda_{bghi}  \\
  &\phantom{=} - 5 \delta_{ac} \lambda_{bdfi} \lambda_{eghi} - \tfrac32\left(2\zeta(3) + \zeta(4)\right) \delta_{cf} \lambda_{adgi} \lambda_{behi}\\
      &\phantom{=} + \tfrac14\delta_{dg} \delta_{eh} \lambda_{acij} \lambda_{bfij} + \delta_{ac} \lambda_{hijk} \left[\tfrac{121}{288} \delta_{bf} \delta_{dg} \lambda_{eijk} -\tfrac{37}{288} \delta_{df} \delta_{eg} \lambda_{bijk}\right]\\
  &\phantom{=} +  \left(\tfrac12-\zeta(3)\right) \delta_{ac} \delta_{bf}  \lambda_{dgij} \lambda_{ehij} + \tfrac18\left(2\zeta(3)-1\right) \delta_{ac} \delta_{bf} \lambda_{deij} \lambda_{ghij}  \\
   &\phantom{=}  + \left(\tfrac5{6} - \zeta(3)\right) \delta_{ac} \delta_{df} \lambda_{beij} \lambda_{ghij} + \left(4 \zeta(3) - 5\right) \delta_{ac} \delta_{df} \lambda_{bgij} \lambda_{ehij}  
  \\ \square_{abc|def|ghi|jkl} &= - 5 \zeta(5) \,\delta_{be} \delta_{fh} \delta_{ik} \delta_{cl} \lambda_{adgj} + \tfrac18 \left(2\zeta(3) -1\right) \delta_{ad} \delta_{be} \delta_{cg} \delta_{fj} \lambda_{hikl}\\
  &\phantom{=}  + \tfrac32 \left(\zeta(4) - 2\zeta(3)\right) \delta_{ad} \delta_{bg} \delta_{eh} \delta_{ij} \lambda_{cfkl} + \tfrac23 \delta_{ad} \delta_{be} \delta_{cg} \delta_{hj} \lambda_{fikl} 
  \end{aligned}
\end{equation}
for later convenience. It turns out that $\bigcirc_{ab|cd} = \bigcirc_{cd|ab}$, which is not necessarily true at higher loop orders.

Utilising the dummy field method \cite{Martin:1993zk,Luo:2002ti,Schienbein:2018fsw}, renormalisation group equations for the scalar cubic couplings $h_{abc}$
 \begin{equation}\label{eq:cubic-gen}
  \begin{aligned}
    \beta_{abc}^{h, 3\ell} &= \left[\gamma_{ae}^{\phi,3\ell} h_{ebc} + \gamma_{be}^{\phi,3\ell} h_{aec} + \gamma_{ce}^{\phi,3\ell} h_{abe} \right] 
     + 12 \zeta(3) \,\lambda_{adij} \lambda_{beig} \lambda_{cfgj} h_{def} \\
    &\phantom{=} - \tfrac12 \left[ \lambda_{aefg} \lambda_{bdfg} \lambda_{ceij} h_{dij} + \text{5 perm.}\right]
    + \tfrac12 \left[ \lambda_{abef} \lambda_{egij} \lambda_{dfij} h_{cdg} + \text{2 perm.}\right] \\
    &\phantom{=} - \tfrac38 \left[ \lambda_{abde} \lambda_{efij} \lambda_{fgij} h_{cdg} + \text{2 perm.}\right]
    - \tfrac12 \left[ \lambda_{abde} \lambda_{cefg} \lambda_{fgij} h_{dij} + \text{2 perm.}\right]\\
    &\phantom{=} - \tfrac12 \left[ \lambda_{adef} \lambda_{deij} \lambda_{bgij} h_{cfg} + \text{2 perm.}\right]
     + 2 \left[ \lambda_{abef}  \lambda_{cdeg} \lambda_{fgij} h_{dij} + \text{2 perm.}\right] \\
    &\phantom{=} + 2 \left[ \lambda_{abef}  \lambda_{fgij} \lambda_{cdij }h_{deg}  + \text{2 perm.}\right] 
     + 2 \left[ \lambda_{bdeg} \lambda_{fgij} \lambda_{cdij} h_{aef}  + \text{5 perm.}\right],\\[0.5em]
     \beta_{abc}^{h, 4\ell} &= \left[\gamma_{ae}^{\phi,4\ell} h_{ebc} +   h_{ade} \lambda_{bcfg} \left(\bigcirc_{de|fg} + \bigcirc_{fg|de}\right) + \text{2 perm.} \right] \\
     &\phantom{=} + \left[ h_{def} \lambda_{abgh} \lambda_{cijk} \left( \bigtriangleup_{gh|def|ijk} + \bigtriangleup_{gh|ijk|def} \right)  + \text{2 perm.} \right] \\
     &\phantom{=} + \left[ h_{ade} \lambda_{bfgh} \lambda_{cijk} \bigtriangleup_{de|fgh|ijk}  + \text{ 5 perm.}  \right] \\
     &\phantom{=} + \left[ h_{def} \lambda_{aghi} \lambda_{bjkl} \lambda_{cmno} \left(\square_{def|ghi|jkl|mno} + \square_{ghi|def|jkl|mno} \right.\right.\\
     &\phantom{=} \left.\left. \qquad\qquad\qquad\qquad\qquad\quad   +\  \square_{ghi|jkl|def|mno} + \square_{ghi|jkl|mno|def}\right) + \text{5 perm.}\right]
  \end{aligned}
\end{equation}
and also mass terms $m^2_{ab}$ 
 \begin{equation}\label{eq:mass-gen}
  \begin{aligned}
    \beta_{ab}^{m^{\!2}\!, 3\ell} &= \left[\gamma_{ae}^{\phi,3\ell} m^2_{eb} + \gamma_{be}^{\phi,3\ell} m^2_{ae}\right] 
    + 6 \zeta(3) \,\lambda_{acde} \lambda_{bcfg} h_{dei} h_{fgi} \\
    &\phantom{=} - \tfrac12 \left[ \lambda_{aefg} \lambda_{bdfg} h_{eij} h_{dij} + \lambda_{acfg} \lambda_{bcij} h_{efg} h_{eij} + \lambda_{acde} \lambda_{bfgi} h_{dei} h_{cfg} \right]\\
     &\phantom{=} + \tfrac12 \lambda_{ceij} \lambda_{dfij} \left[ h_{acd} h_{bef} +  \lambda_{abcd} \,  m^2_{ef}\right] - \tfrac38 \lambda_{dfij} \lambda_{efij} \left[ h_{acd} h_{bce} + \lambda_{abcd} \,m^2_{ce}\right] \\
     &\phantom{=} - \tfrac12 \lambda_{fgij} \!\left[\lambda_{acfg} \lambda_{bdij} \, m^2_{cd} + \tfrac12 \lambda_{abcd} h_{cfg} h_{dij} + \lambda_{acgh} h_{bcd} h_{dij} + \lambda_{bcgh} h_{acd} h_{dij}\right]\\
    &\phantom{=} + 2\lambda_{defg} \!\left[\lambda_{acei} (\lambda_{bfgi}  m^2_{cd} \!+\! h_{fgi} h_{bcd})  + (\tfrac12\lambda_{abcd}  h_{fgi} \!+\! \lambda_{afgi} h_{bcd})h_{cei} + (a \!\leftrightarrow\! b) \right], \\[0.5em]
    \beta_{ab}^{m^{\!2}\!, 4\ell} &= \left[\gamma_{ae}^{\phi,4\ell} m^2_{eb} + \gamma_{be}^{\phi,4\ell} m^2_{ae}\right] \\
    &\phantom{=} + m^2_{cd} \left[ \lambda_{abef} \left( \bigcirc_{cd|ef} + \bigcirc_{ef|cd}\right)  + \lambda_{aefg} \lambda_{bhij} \left( \bigtriangleup_{cd|efg|hij} + \bigtriangleup_{cd|hij|efg}  \right) \right] \\
    &\phantom{=} + h_{acd} h_{bef} \left[ \bigcirc_{cd|ef} + \bigcirc_{ef|cd} \right] +  h_{cde} h_{fgh}  \lambda_{abij}  \ \bigtriangleup_{ij|cde|fgh}  \\
    &\phantom{=} + \left[h_{aef} h_{ghi} \lambda_{bjkl}  \left( \bigtriangleup_{ef|ghi|jkl} + \bigtriangleup_{ef|jkl|ghi}\right) + (a\leftrightarrow b)\right]\\
    &\phantom{=} + \left[h_{cde} h_{fgh} \lambda_{aijk} \lambda_{blmn} \left(\square_{ijk|lmn|cde|fgh} + \square_{ijk|cde|lmn|fgh} +  \square_{ijk|cde|fgh|lmn}\right.\right.\\
    &\phantom{=}  \left.\left. \qquad\qquad\   + \ \square_{cde|ijk|lmn|fgh}  + \square_{cde|ijk|fgh|lmn} + \square_{cde|fgh|ijk|lmn} \right) + (a\leftrightarrow b)\right]
  \end{aligned}
\end{equation}
of any scalar sector are determined. Once again, the abbreviations `$p \text{ perm.}$' denote $p$ swaps of external indices producing inequivalent expressions.
Moreover, \eq{WF-gen} is also a part of the three-loop Yukawa $\beta$-function.
The equations \eq{WF-gen}, \eq{quartic-gen}, \eq{cubic-gen} and \eq{mass-gen} represent a partial result of the complete three- and four-loop expressions for the general renormalisable QFT \eq{master-template}.

Our results extracted from \eq{THDM-RGE} have been cross-checked against literature expressions. 
The general expressions are in agreement with \cite{Jack:1990eb, Jack:2018oec}.
In $O(n)$ theories \eq{O(n)}, three and four-loop quartic $\beta$ functions, mass and field anomalous dimensions \cite{Brezin:1974eb,Dittes:1977aq,Kazakov:1979ik,Chetyrkin:1981jq,Gorishnii:1983gp,Kazakov:1984km,Kleinert:1991rg,Batkovich:2016jus,Schnetz:2016fhy,Kompaniets:2017yct} are reproduced. The same holds also for $O(n)\times O(m)$ theories \cite{Pelissetto:2001fi, Calabrese:2003ww, Kompaniets:2019xez} and models with a cubic anisotropy \cite{Nelson:1974xnq, Kleinert:1994td, Adzhemyan:2019gvv}.
Moreover, three-loop RGEs for the full THDM scalar potential, featuring 10 real quartics and 4 mass parameters, are in agreement with \cite{Bednyakov:2018cmx}. Finally, the results have been checked against a theory with a single, real scalar field $\phi$, which trivially amounts to dropping all indices in \eq{WF-gen} -- \eq{mass-gen}.

Furthermore, completely new results can also be obtained using these expressions. 
This is demonstrated by computing RG equations for a $U(n) \times U(n)$ scalar theory, with the complex matrix field $\phi_{ab}$ at three-loop order. The bare action is given by 
\begin{equation}
  \mathcal{L} = \text{tr} \left[ \partial_\mu \phi^\dagger \partial^\mu \phi\right] - m^2 \text{tr} \left[ \phi^\dagger  \phi\right] - v\, \text{tr} \left[ \phi^\dagger  \phi \right] \text{tr} \left[ \phi^\dagger  \phi \right] - u \, \text{tr} \left[ \phi^\dagger  \phi \phi^\dagger  \phi\right],
\end{equation}
and is distinct from that of a $O(n) \times O(2n)$ theory due to the single-trace quartic interaction.
In the large-$n$ limit, two-loop results have been presented in \cite{Bond:2017tbw}, and can be extended for finite-$n$ using \cite{Machacek:1983tz,Machacek:1983fi,Machacek:1984zw,Luo:2002ti,Schienbein:2018fsw,Poole:2019kcm}.
For convenience, we introduce the 't Hooft couplings of the quartic sector
\begin{equation}
  \alpha_u = n\,u/(4\pi)^2 \qquad\text{and}\qquad \alpha_v = n^2\,v/(4\pi)^2.
\end{equation}
Using \eq{WF-gen}, the field anomalous dimension reads up to three-loop order
\begin{equation}
\begin{aligned}\label{eq:U(n)xU(n)-GammaPhi}
    \gamma_{\phi} = &\  2\left[1 + \tfrac1{n^{2}}\right] \alpha_u^2   +\tfrac8{n^{2}} \alpha_u \alpha_v +  \tfrac{2}{n^{2}} \left[ 1 + \tfrac{1}{n^{2}}\right] \alpha_v^2\\
  & - 4 \left[1 + \tfrac4{n^2}\right] \alpha_u^3 - \tfrac6{n^2} \left[7 + \tfrac3{n^2}\right]\alpha_u^2 \alpha_v - \tfrac{12}{n^2} \left[1 + \tfrac4{n^2}\right]\alpha_u  \alpha_v^2 - \tfrac2{n^2} \left[1 + \tfrac5{n^2}  + \tfrac4{n^4}\right] \alpha_v^3
\end{aligned}
\end{equation}
From \eq{mass-gen}, the mass anomalous dimension $\gamma_{m^2} = m^{-2} \, \beta_{m^2}$  
\begin{equation}
  \begin{aligned}
  \gamma_{m^2} = &\ 8\,\alpha_u + 4 \left[1 + \tfrac1{n^2}\right] \alpha_v   - 20\left[1 + \tfrac1{n^2}\right] \alpha_u^2 - \tfrac{80}{n^2} \alpha_u \alpha_v - \tfrac{20}{n^2}\left[1 + \tfrac{1}{n^2}\right] \alpha_v^2 \ + \\
  &  24 \!\left[ 10 \!+\! \tfrac{37}{ n^2}\right]\! \alpha_u^3 + 12 \!\left[1\! +\! \tfrac{199}{n^2} \!+ \!\tfrac{82}{n^4}\right]\! \alpha_u^2 \alpha_v + \tfrac{72}{n^2} \!\left[ 10 \!+ \!\tfrac{37}{n^2}\right] \!\alpha_u \alpha_v^2 + \tfrac{12}{n^2}\!\left[10 \!+ \!\tfrac{47}{ n^2} \! + \!\tfrac{37}{n^4}\right] \!\alpha_v^3
  \end{aligned}
\end{equation}
is obtained. Finally, \eq{quartic-gen} yields $\beta$ functions for the 't Hooft couplings $\alpha_{u,v}$ 
\begin{equation}
  \begin{aligned}
  \beta_{v} = &\ 12\, \alpha_u^2 + 16\,\alpha_u \alpha_v + 4 \left[1 + \tfrac4{n^2}\right] \alpha_v^2 - 96\,\alpha_u^3 - 40 \left[1 + \tfrac{41}{5 n^2}\right] \alpha_u^2 \alpha_v - \tfrac{352}{n^2} \alpha_u \alpha_v^2 \\
  & - \tfrac{24}{n^2 } \left[3 + \tfrac7{n^2}\right] \alpha_v^3 + \left[ 772 + 384 \zeta(3) + \tfrac{1700 + 1536 \zeta(3)}{n^2}\right] \alpha_u^4  \\
  & + 96 \left[ 5 + 4\tfrac{25 + 12 \zeta(3)}{n^2}\right] \alpha_u^3 \alpha_v + \left[12 + 8 \tfrac{835 + 144 \zeta(3)}{n^2} + 36 \tfrac{291 + 224 \zeta(3)}{n^4}\right] \alpha_u^2 \alpha_v^2  \\
  & + \tfrac{16}{n^2} \left[79 + \tfrac{659 + 384 \zeta(3)}{n^2} \right] \alpha_u \alpha_v^3 + \tfrac4{n^2} \left[33 + \tfrac{461 + 240 \zeta(3)}{n^2} + \tfrac{740 + 528 \zeta(3)}{n^4}\right] \alpha_v^4,
  \end{aligned}
\end{equation}
\begin{equation}\label{eq:betaV}
  \begin{aligned}
    \beta_{u} = &\ 8\,\alpha_u^2 + \tfrac{24}{n^2} \alpha_u \alpha_v   - 24\left[1 + \tfrac{5}{n^2}\right] \alpha_u^3 - \tfrac{352}{n^2} \alpha_u^2 \alpha_v - \tfrac{8}{n^2}\left[5 + \tfrac{41}{n^2}\right] \alpha_u \alpha_v^2 \\
  & + 104 \left[ 1 + \tfrac{295 + 144 \zeta(3)}{13 n^2}\right] \alpha_u^4 + \tfrac{32}{n^2}\left[91 + 48 \zeta(3) + \tfrac{211 + 192 \zeta(3)}{n^2}\right] \alpha_u^3 \alpha_v \\
  &- \tfrac8{n^2} \left[35 - \tfrac{1591 + 1152\zeta(3)}{n^2}\right] \alpha_u^2 \alpha_v^2 - \tfrac8{n^2}\left[ 13 - \tfrac{184 + 96 \zeta(3)}{n^2}  - \tfrac{821+ 672 \zeta(3)}{n^4}\right] \alpha_u \alpha_v^3.\\
  \end{aligned}
\end{equation}
At four-loop order, the scalar field anomalous dimension receives the contribution
\begin{equation}
  \begin{aligned}
    \gamma_\phi^{4\ell} = &\ 10 \left[1 + \tfrac{45}{n^2} + \tfrac{20}{n^4}\right] \alpha_u^4 + \tfrac{80}{n^2}\left[ 7 + \tfrac{26}{n^2}\right] \alpha_u^3 \alpha_v - \tfrac{60}{n^2}\left[2 - \tfrac{51}{n^2} -  \tfrac{17}{n^4}\right] \alpha_u^2 \alpha_v^2 \\
    & - \tfrac{80}{n^2} \left[ 1 - \tfrac{9}{n^2} - \tfrac{25}{n^4} \right] \alpha_u \alpha_v^3 - \tfrac{10}{n^2} \left[ 1 - \tfrac8{n^2} - \tfrac{34}{n^4} - \tfrac{25}{n^6}\right] \alpha_v^4.
  \end{aligned}
\end{equation}
We obtain corrections for the mass anomalous dimension
\begin{equation}
  \begin{aligned}
    \gamma_{m^2}^{4\ell} = &\ - 128 \left[ \tfrac{1501}{96} + \zeta(3) + 3 \zeta(4)  + \tfrac{5375}{32 n^2} + \tfrac{43 \zeta(3)}{2 n^2} + \tfrac{33\zeta(4)}{n^2} \right] \alpha_u^4 \\
    & - \tfrac{64}{3} \left[ 25 - 12 \zeta(3) + \tfrac{3687}{2n^2} + \tfrac{72}{n^2}\left(4 \zeta(3) + 5 \zeta(4)\right) \right] \alpha_u^3 \alpha_v \\
    & - \tfrac{144}{n^2}\left[51 + 24 \zeta(3) + 8 \zeta(4)\right] \alpha_u^2 \alpha_v^2 + \tfrac{4}{n^2} \left[\tfrac13 - 48 \zeta(3)\right] \left(8 \alpha_u + \alpha_v \right) \alpha_v^3,
  \end{aligned}
\end{equation}
as well as the single and double trace quartic interactions
 \begin{equation}
   \begin{aligned}
     \beta_u^{4\ell} = &\,- 8\left[67 - 32 \zeta(3) + 80 \zeta(5) + \tfrac{4543}{n^2}  + \tfrac{3392 \zeta(3)}{n^2} - \tfrac{864 \zeta(4)}{n^2}  + \tfrac{3760 \zeta(5)}{n^2} \right] \alpha_u^5 \\
     & - \tfrac{128}{n^2} \left[249 + 190 \zeta(3) - 72 \zeta(4) + 200 \zeta(5) \right] \alpha_u^4 \alpha_v - \tfrac{256}{n^2} \left[ \tfrac{10}3  - 9 \zeta(3) \right] \alpha_u^2 \alpha_v^3 \\
     & + \tfrac{384}{n^2} \left[ 5 + 2 \zeta(3) + 4 \zeta(4)\right] \alpha_u^3 \alpha_v^2  - \tfrac8{n^2} \left[29 - 48 \zeta(3)\right] \alpha_u \alpha_v^4, 
     \end{aligned}
 \end{equation}
 and 
  \begin{equation}
   \begin{aligned}
     \beta_v^{4\ell} = &\, -\! 256 \left[ \tfrac{97}{6} + 25 \zeta(3) - 6 \zeta(4) + 15 \zeta(5)   + \tfrac{2515}{12n^2} + \tfrac{225\zeta(3) }{n^2} - \tfrac{51 \zeta(4) }{n^2} + \tfrac{340 \zeta(5)}{n^2}   \right] \alpha_u^5\\
     &- 256 \left[ \tfrac{1501}{96} + \zeta(3) + 3 \zeta(4) + \tfrac{18483}{32 n^2} + \tfrac{562 \zeta(3)}{n^2} - \tfrac{93 \zeta(4)}{n^2} + \tfrac{580 \zeta(5)}{n^2} \right] \alpha_u^4 \alpha_v \\
     & - 256 \left[ \tfrac{25}{12} - \zeta(3) + \tfrac{1235}{4n^2}  + \tfrac{211 \zeta(3)}{n^2} + \tfrac{6 \zeta(4)}{n^2} + \tfrac{80 \zeta(5)}{n^2} \right] \alpha_u^3 \alpha_v^2 \\
     & - \tfrac{768}{n^2} \left[ \tfrac{45}{8} + 14 \zeta(3) \right] \alpha_u^2 \alpha_v^3  + \tfrac{256}{n^2} \left[ \tfrac73 - 3 \zeta(3) \right] \alpha_u \alpha_v^4 + \tfrac{40}{3n^2} \alpha_v^5,
     \end{aligned}
 \end{equation}
where only leading and next-to-leading terms in $n^{-2}$ are included in the interest of brevity.\footnote{An auxiliary file listing the full expressions is provided.}

In summary, we have shown how fully general RGEs can be extracted from expressions obtained in much simpler models. This represents an opportunity to break involved higher-loop computations apart into smaller problems. In doing so, existing results can be reused. 
 We have explored this technique for purely scalar RGEs as far as possible, and obtained three-loop results. Extracting general four-loop RGEs requires more input data.
 We have refrained from using Weyl consistency conditions as in \cite{Jack:2018oec} to provide an independent cross-check.
  
In principle, the idea can be used to extract other terms and/or different RGEs, possibly feeding off the data \cite{Bednyakov:2012en,Chetyrkin:2012rz,Bednyakov:2013eba,Chetyrkin:2013wya,Bednyakov:2013cpa,Bednyakov:2014pia,Davies:2019onf}, and supplemented Weyl consistency conditions \cite{Poole:2019kcm}. Explorations into this direction are left for future work. However, such an extension requires a careful treatment of terms that are sensitive to the $\gamma_5$-problem as soon as fermions are involved, see for instance \cite{Zoller:2015tha,Bednyakov:2015ooa,Poole:2019txl,Poole:2019kcm}. This is because the dimensional regularisation scheme is incompatible with chiral symmetry. Moreover, supersymmetry is inherently violated, which motivates a change of scheme to facilitate a similar procedure.

\section*{Acknowledgements}
The author is thankful to Mark Steudtner, Dominik Stöckinger and Jack Setford for comments on the draft, Hugh Osborn, Florian Herren and Daniel Litim for pointing out results in the literature as well as Charanjit Kaur Khosa for helpful discussions. Furthermore, TS is indebted to Alexander Bednyakov for cross-checking parts of the results and pointing out mistakes. 

\bibliographystyle{JHEP}
\bibliography{ref}
\end{document}